 \documentclass[final,5p,times,twocolumn,authoryear]{elsarticle}

\usepackage{amssymb}
\usepackage{lipsum}
\usepackage{hyperref}
\usepackage{multirow}
\usepackage{amsmath}
\usepackage{graphicx}
\usepackage{float}
\usepackage{placeins}

\journal{High Energy Astrophysics}

\begin{document}

\begin{frontmatter}



\title{Potential detection of $\sim$ 4.2 keV emission line from GRS 1747–312}

\author[aff1]{Amom Lanchenbi Chanu}
\author[aff2]{Akash Garg}
\author[aff2]{Ranjeev Misra}
\author[aff1]{A.~Senorita Devi}

\affiliation[aff1]{organization={Department of Physics, Manipur University},
	addressline={Canchipur},
	city={Imphal West},
	postcode={795003},
	state={Manipur},
	country={India}}

\affiliation[aff2]{organization={Inter-University Centre for Astronomy and Astrophysics},
	addressline={Ganeshkhind},
	city={Pune},
	postcode={411007},
	state={Maharashtra},
	country={India}}

\begin{abstract}
We present a broadband spectral analysis of the neutron star LMXB GRS 1747-312 using $\sim 40$ ks {\it Astrosat} data. The source was observed during the decay phase of the  2017 outburst, with an absorbed 1.0-5.5 keV flux of $1.67^{+0.04}_{-0.07} \times 10^{-11}$ erg s$^{-1}$ cm$^{-2}$, corresponding to a luminosity of $\sim (0.9-1.80) \times 10^{35}$ erg s$^{-1}$. The continuum is modeled with thermal Comptonization of blackbody emission and interstellar absorption. A mildly broad iron line at $\sim 6.4$ keV is fitted with a disc reflection component. Narrow lines below 2 keV are described by a hot plasma using the \textsc{xspec} model \texttt{apec}. Additionally, there is a potential detection of an emission line at $4.19^{+0.12}_{-0.10}$ keV with width $\sigma = 0.2 \pm 0.2~\mathrm{keV}$ and line flux = $13^{+10}_{-9} \times 10^{-5}$ erg s$^{-1}$ cm$^{-2}$. Examination of several short duration ($\sim$ few kiloseconds) {\it Swift} observations at few times the {\it AstroSat} source flux, provided  upper limits to the line flux $< 30 \times 10^{-5}$ erg s$^{-1}$ cm$^{-2}$. The 4.2 keV line likely originates from reflection off the neutron star surface. Shifting the neutral Fe $K_\alpha$ line from its rest energy of 6.4 to 4.2 keV requires a redshift of $z \sim 0.6$, consistent with that expected from the surface of a non-spinning $ 1.4 M_\odot$, 10 km radius neutron star. If confirmed, this feature provides a potential direct measurement of gravitational redshift, allowing us to place strong constraints on the neutron star's mass-to-radius ratio and gain valuable insights into the equation of state (EOS) of dense-matter.

\end{abstract}



\begin{keyword}
X-rays: binaries \sep galaxies: nuclei \sep stars: neutron stars \sep accretion, accretion discs



\end{keyword}

\end{frontmatter}




\section{Introduction}
\label{introduction}

GRS 1747-312 is a neutron star low-mass X-ray binary \citep[LMXBs; see reviews, for instance, on LMXBs in general and NS LMXBs in particular, respectively][]{Bahramian2024,disalvo2023review} located in the metal-rich globular cluster Terzan 6 at a distance of  $9.5^{+3.3}_{-2.5}$ kpc \citep{Barbuy1997ntt, hulleman2003bursts, kuulkers2003photospheric, in2004bepposax}. It was discovered in 1990 by {\it ROSAT} and {\it GRANAT} \citep{predehl1991rosat, pavlinsky1994x} and confirmed as a neutron star through type-I X-ray bursts \citep{iwai2017broad, hulleman2003bursts}. The system exhibits eclipses with a 12.36 hr orbital period and undergoes regular outbursts every 4-5 months \citep{bazzano2000orbital}. During outburst, it reaches X-ray luminosity of $\sim10^{36}$ erg s$^{-1}$, while in quiescence it drops to $\sim10^{33}$ erg s$^{-1}$ \citep{saji2016peculiar, vats2018quiescent, painter2024possible}. As one of only 13 Galactic LMXBs with sharp eclipses during outburst \citep{Bahramian2024}, it remains a compelling target for study.

While extensively studied in outbursts, the quiescent phase of GRS 1747-312 reveals puzzling features. A $\sim$ 4.2 ks eclipse was observed in a {\it Chandra} 2021 quiescent observation (ObsID 23443), longer than typical outburst eclipses \citep{van2024discovery}. This suggests possible contamination from nearby sources, as GRS 1747-312 becomes faint in quiescence. Its position, first identified with {\it Chandra} HRC-I during outburst \citep{revnivtsev2002localization}, was later refined by \cite{hulleman2003bursts}. In the 2009 {\it Suzaku} observation, \cite{saji2016peculiar} reported emission from the cluster core in the absence of outburst, implying a possible unrelated source. More recently, \cite{van2024discovery} identified a nearby source, Terzan 6 X-2, only $\sim$ 0.7$^{\prime\prime}$ away, as the dominant contributor during quiescence.

Several models have been proposed to explain the unusual quiescent behavior and eclipse features of GRS 1747-312. \cite{painter2024possible} suggested that the system could be part of a hierarchical triplet, in which a third outer companion which may possibly be a brown dwarf, M-dwarf, or even a massive planet orbiting the central low-mass X-ray binary (LMXB). Another possibility is that two closely spaced LMXBs exist within the globular cluster, potentially forming a gravitationally bound pair. \cite{saji2016peculiar} proposed that the emission detected during the eclipsing or off states may arise from an unresolved interloper source near the primary system. Alternative explanations include residual emission arising from a vertically extended corona that remains unobscured by the companion star. Another possibility is modulation of the observed emission geometry caused by a precessing inner accretion disk, rather than intrinsic changes in the overall accretion structure. 

Spectral studies during outburst typically involve a blackbody and Comptonized component \citep{intzand2005outburst}, however the quiescent spectrum is less understood. While no significant line features were detected during the outburst in {\it BeppoSAX} observations \citep{in2004bepposax, saji2016peculiar}, a narrow 6.4 keV Fe $K_\alpha$ line was identified in quiescence with {\it Chandra}, which may suggest reflection from a precessing accretion disk \citep{painter2024possible}.

In this study, we present a broadband spectral analysis of GRS 1747-312 using a \texorpdfstring{$\sim 40$}{~40} ks {\it AstroSat} observation obtained during  the decay phase of the 2017 outburst. We also analyze non-simultaneous observations of the source taken with {\it Chandra}, {\it Swift}, and {\it XMM-Newton} during epochs outside the major outburst phases, spanning quiescent and low-luminosity accretion states. The paper is organized as follows: Section \ref{sec:obs} describes the observations and data reduction; Section \ref{sec:results} presents the spectral analysis and results; and Section \ref{sec:discussion} discusses the findings and provides the conclusions.

\section{Observations \texorpdfstring{\&}{&} Data Reduction}
\label{sec:obs}
In the present work, we used an {\it Astrosat} archival dataset for GRS 1747-312, taken on July 22$^{nd}$ and 23$^{rd}$, 2017 (Observation ID: G07$\_020T01\_9000001398$). Eight {\it Swift}/XRT observations, five  {\it  Chandra} observations, and one {\it XMM-Newton} observation were also used in the study. The details of the observations are given in Table \ref{tab:obs}.
\subsection{Astrosat} 

{\it AstroSat}, India’s first multi-wavelength astronomy satellite, enables broadband X-ray studies (0.5-80 keV) of X-ray binaries using two of its co-aligned instruments: the Large Area X-ray Proportional Counter (LAXPC) and the  Soft X-ray Telescope (SXT) \citep{agrawal2006broad}. LAXPC consists of three units (LAXPC10, 20 and 30), each with 3–80 keV coverage, 10 $\mu$s timing resolution, and a combined effective area of $\sim$ 6000 cm$^2$ \citep{yadav2016astrosat, agrawal2017large, antia2017calibration}. Due to the gain instability in LAXPC30 \citep{antia2017calibration}, only LAXPC10 and LAXPC20 data were used.

Data reduction follows standard procedures using the {\bf{\it LAXPC} software22Aug15}\footnote{\url{http://astrosat-ssc.iucaa.in/laxpcData}}. Level 2 event files are generated from Level 1 using {\tt laxpc\_make\_event}, and Good Time Interval (GTI) files are produced with {\tt laxpc\_make\_stdgti} to exclude periods affected by the South Atlantic Anomaly (SAA) and Earth occultation effects\citep{agrawal2017large}. Given the source faintness, lightcurves and spectra are extracted from the top layer (L1) of LAXPC10 and 20 using {\tt laxpc\_make\_lightcurve} and {\tt laxpc\_make\_spectra} to minimize background \citep{antia2021large, beri2019thermonuclear, nath2022astrosat}. Corresponding background and response files are used from the ASSC repository.

The SXT operates in the 0.5-7.0 keV range, with an effective area of $\sim$ 90 cm$^2$ at 1.5 keV, and is optimized for medium-resolution spectroscopy \citep{singh2017soft}. Cleaned Level 2 event files are merged using {\tt SXTMerger}\footnote{\url{http://astrosat-ssc.iucaa.in/sxtData}}. Source spectra are extracted via \textsc{XSELECT} from a 15$^\prime$ circular region. Pile-up correction is unnecessary as the count rate is below 40 counts s$^{-1}$. The standard {\tt RMF (sxt\_pc\_mat\_g0to12.rmf)} is used, and vignetting-corrected {\tt ARFs} are generated via {\tt sxtARFModule}. We used \textsf{SkyBkg\_sxt\_LE0p35\_R16p0\_v05\_Gd0to12.pha} as the background file for the SXT spectrum. The final spectra are grouped and binned using {\tt grppha} and {\tt FTGROUPPHA} \citep{kaastra2016optimal}.

\subsection{XMM-Newton} 

The \textit{XMM-Newton} EPIC-pn operates over the 0.15-15 keV energy range, while EPIC-MOS operates over 0.2-12 keV, with spectral resolutions of approximately 80 eV and 70 eV at 1 keV, respectively. We used a 2004 observation (ObsID: 0206990101) with a 14 ks exposure. Data from the EPIC-PN and MOS detectors were processed using the \texttt{epproc} and \texttt{emproc}\footnote{\url{https://xmm-tools.cosmos.esa.int/external/sas/current/doc/epproc/index.html}} tools in SAS v20.0 \citep{struder2001european, turner2001european}. Standard filtering criteria were applied: \texttt{\#XMMEA\_EP}, \texttt{PATTERN}~$\leq 4$, \texttt{FLAG}=0 for PN; and \texttt{\#XMMEA\_EM}, \texttt{PATTERN}~$\leq 12$ for MOS. Good time intervals were selected by excluding high background flaring periods based on count-rate filtering. The source was identified using the \texttt{edetectchain} task. Source spectra were extracted from a 30$^{\prime\prime}$ circular region centered on the source, and background spectra from a nearby source-free region of equal size. Response files were generated using \texttt{rmfgen} and \texttt{arfgen}.

\subsection{Chandra} 
{\it Chandra} observational data from five quiescent-state observations conducted in 2021 and 2023, using the ACIS-S instrument in Continuous Clocking (CC) mode and covering the 0.3–10 keV energy range, were analyzed with HEASoft 6.33.1 and CIAO 4.16. X-ray point sources were identified using {\tt\string wavdetect}\footnote{\url{https://cxc.cfa.harvard.edu/ciao/ahelp/wavdetect.html}}. Spectral products were generated from the Level 2 event file using {\tt specextract}, which executes {\tt dmextract}, {\tt mkarf}, {\tt arfcorr}, {\tt mkrmf}, {\tt dmgroup}, and {\tt dmhedit} to produce and calibrate the spectra, ARF, and RMF files. A 10$^{\prime\prime}$ circular region centered on the source coordinates was used for source extraction, while background spectra were obtained from a nearby source-free region of the same size. The resulting spectra were rebinned to a minimum of 30 counts per bin using {\tt grppha}.

\subsection{Swift X-ray Observatory}  
The {\tt Swift}/XRT, with an effective area of $\sim$ 110 cm$^2$ at 1.5 keV and operating over 0.2-10 keV, observed the source in {\tt Photon Counting (PC)} mode. To ensure data quality, only observations with off-axis angles $<$ 3$'$ and exposure times $>$ 900 s were included in the analysis.

Data were processed using the {\tt Automated XRT Analysis Tools}\footnote{\url{https://www.swift.ac.uk/user_objects/docs.php}}, with {\tt xrtpipeline} applying standard screening (grades 0-12 for PC mode). Source and background spectra were extracted using circular regions of 0.8$^{\prime\prime}$ and 1.6$^{\prime\prime}$ in radius, respectively, with adjustments applied when necessary to mitigate pile-up or contamination. Ancillary response files (ARFs) were generated using {\tt xrtmkarf}, and spectral products were created using {\tt xrtmkresp}. Only observations with net source count rates $>$ 0.1 counts s$^{-1}$ were included.

Final spectra for {\tt Chandra}, {\tt Swift}, and {\tt XMM-Newton} were grouped using {\tt grppha} with a minimum of 30 counts per bin to enable $\chi^2$ statistics and preserve spectral resolution.

\setlength{\tabcolsep}{3pt}

\begin{table*}
	\centering
	\caption{Observation log of GRS 1747$-$312}
	\label{tab:obs}
	\resizebox{\textwidth}{!}{%
		\begin{tabular}{lcccccc}
			\hline
			Telescope & Observation ID & Date (YYYY-MM-DD) & Instrument & Exposure (ks) & MJD$_{\mathrm{start}}$ & MJD$_{\mathrm{stop}}$ \\
			\hline
			\textit{AstroSat} & G07\_020T01\_9000001398 & 2017-07-22 & SXT, LAXPC10, LAXPC20 & 41 & 57956.67403 & 57957.95707 \\
			\textit{Chandra}  & 23443 & 2021-04-20 & ACIS-S & 30 & 59324.8420 & 59325.2131 \\
			& 23441 & 2021-08-23 & ACIS-S & 10 & 59449.4127 & 59449.5506 \\
			& 26513 & 2023-08-17 & ACIS-S & 22 & 60173.0153 & 60173.2895 \\
			& 26514 & 2023-08-27 & ACIS-S & 20 & 60183.9716 & 60184.2222 \\
			& 26515 & 2023-09-07 & ACIS-S & 22 & 60194.9736 & 60195.2534 \\
			\textit{Swift} & 00032761001 & 2013-03-18 & XRT & 0.9 & 56369.8766 & 56369.8882 \\
			& 00032761002 & 2013-03-24 & XRT & 0.9 & 56375.0934 & 56375.1049 \\
			& 00032761004 & 2016-05-19 & XRT & 1.4 & 57527.6441 & 57527.6614 \\
			& 00032761007 & 2016-05-20 & XRT & 1.4 & 57528.6400 & 57528.6570 \\
			& 00032761008 & 2017-10-09 & XRT & 0.9 & 58035.5862 & 58035.5978 \\
			& 00032761009 & 2021-04-05 & XRT & 0.9 & 58037.44153 & 58037.46182 \\
			& 00032761023 & 2021-08-07 & XRT & 1.2 & 58643.39931 & 58643.41115 \\
			& 00032761039 & 2021-06-21 & XRT & 0.9 & 59391.36524 & 59391.37572 \\
			\textit{XMM-Newton} & 0206990101 & 2004-09-28 & PN, MOS1, MOS2 & 14.6 & 53276.62325 & 53276.71841 \\
			\hline
		\end{tabular}%
	}
\end{table*}

\begin{figure}[h]
	\centering
	\includegraphics[width=0.35\textwidth,angle=-90]{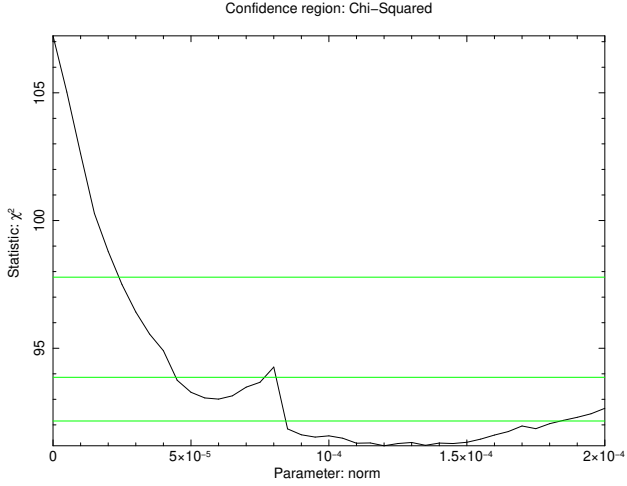}
	\caption{$\chi^{2}$ profile obtained using the XSPEC {\tt steppar} command by stepping the normalization of the Gaussian component associated with the $\sim$4.2 keV candidate emission line. The horizontal lines correspond to the 68$\%$, 90$\%$, and 99$\%$ confidence levels ($\Delta \chi^{2}$ = 1.0, 2.71, and 6.63) for normalization, showing that zero normalization is excluded and supporting the presence of the candidate emission feature.}
	\label{fig:steppar}
\end{figure}

\begin{figure*}
	\centering
	
	\begin{minipage}[b]{0.48\textwidth}
		\centering
		\includegraphics[width=0.75\linewidth, angle=-90]{figure1a.eps}
		\\[8pt] (a)
	\end{minipage}
	\begin{minipage}[b]{0.48\textwidth}
		\centering
		\includegraphics[width=0.75\linewidth, angle=-90]{figure1b.eps}
		\\[8pt] (b)
	\end{minipage}
	
	
	\begin{minipage}[b]{0.48\textwidth}
		\centering
		\includegraphics[width=0.75\textwidth, angle=-90]{figure1c.eps}
		\\[8pt] (c)
	\end{minipage}
	\begin{minipage}[b]{0.48\textwidth}
		\centering
		\includegraphics[width=0.75\textwidth, angle=-90]{figure1d.eps}
		\\[8pt] (d)
	\end{minipage}
	
	\caption{Unfolded energy spectra in 1.0-50 keV as observed by SXT (black), LAXPC20 (red), and LAXPC10 (green). Panels (a)-(d) show the fitted spectra with:(a) \texttt{constant*tbabs*(thcomp*bbodyrad+gaussian)},
		(b) \texttt{constant*tbabs*(thcomp*bbodyrad+gaussian+apec)}, (c) \texttt{constant*tbabs*(thcomp*bbodyrad+gaussian+gaussian+apec)}, (d) \texttt{constant*tbabs*(thcomp*bbodyrad+xillverCp+relxillCp+apec)}. Each panel shows the best-fit model (top) and residuals (bottom).}
	\label{fig:1}
\end{figure*}

\section{Results}
\label{sec:results}
\subsection{Astrosat} 
We use SXT (1-5.5 keV) and LAXPC (4-50 keV) spectra to minimize background contamination and enable reliable broadband spectral fitting across the 1-50 keV energy range.  A systematic uncertainty of 3\% was applied to both the SXT and LAXPC spectra during spectral fitting to account for calibration uncertainties \citep{misra2017astrosat}. The combined spectrum is initially fitted using the model
\texttt{constant*tbabs*(thcomp*bbodyrad)}, which represents thermal Comptonization of seed photons originating from the neutron star’s {\bf surface or boundary layer}. The {\tt constant} component was included to account for cross-calibration differences between instruments, fixed at unity for the SXT and allowed to vary for the LAXPC detectors. The \texttt{tbabs} component models interstellar absorption along the line of sight \citep{wilms2000absorption}. The \texttt {thcomp} model \citep{zdziarski2020spectral} describes thermal Comptonization by a hot electron population, and is characterized by the photon index ($\Gamma_\tau$), electron temperature ($kT_e$), covering fraction ($\mathrm{cov}_{\mathrm{frac}}$), and redshift ($z$) of the source. Due to poor constraints, $kT_e$ is fixed at 100 keV. The \texttt{bbodyrad} component models blackbody emission from the neutron star surface, with normalization given by norm$_{\text{bbody}} = (R_{\text{km}} / D_{10})^2$. As {\tt thcomp} is a convolution model, the energy grid is extended to 0.01-200 keV using 2000 logarithmic bins. This fit yields poor statistics ($\chi^2$/d.o.f. = 322.1/128) and the fitting does not improve if the photon source is considered to be a disc rather than a blackbody. Including a Gaussian emission line at 6.4 keV improved the fit ($\chi^2$/d.o.f. = 140.8/125) considerably. The corresponding spectral fit and residuals are shown in Figure \ref{fig:1}(a), where narrow emission features below 2 keV are evident in the residuals. This motivated the addition of an optically thin plasma model ({\tt apec}) with temperature, abundance, and normalization left as free parameters, resulting in a substantial improvement in the fit  ($\chi^2$/d.o.f. = 106.3/122). To assess the energy calibration of SXT, we allowed the `redshift' parameter of the \texttt{apec} component to vary. The resulting value is consistent with zero and yields an upper limit of $z < 1.4 \times 10^{-2}$ (at the 2$\sigma$ confidence level), indicating that the SXT energy calibration is adequate for identifying emission-line features. The best fit spectrum for this model and residuals are shown in Figure \ref{fig:1}(b). There is a potential feature at $\sim 4.2$ keV, and to address that, we included a second Gaussian, which resulted in  $\chi^2$/d.o.f. = 92.9/119. The best fit line energy was found to be $4.19^{+0.12}_{-0.10}$ keV with width $\sigma = 0.2 \pm 0.2~\mathrm{keV}$ and line flux $=13^{+10}_{-9} \times 10^{-5}$ erg s$^{-1}$cm$^{-2}$. This spectral fitting is shown in Figure \ref{fig:1}(c).

To assess the statistical significance of the $\sim$ 4.2 keV emission feature, we performed several tests. The ratio of the Gaussian line normalization to its $1\sigma$ uncertainty gives $K/\Delta K \approx 3.05 $, corresponding to an approximate $\sim 3\sigma$ detection. We further performed Monte Carlo simulations using the {\tt simftest} command in XSPEC with 1000 simulated spectra, yielding an average null-hypothesis probability of 0.003. The variation of $\chi^{2}$ as a function of the Gaussian normalization obtained using the {\tt steppar} command is shown in Figure 1, confirming the potential presence of the emission feature. 

The Gaussian emission line at $\sim 6.4$ keV should be the Iron fluorescence line from the disk, which should be accompanied by a reflection bump. To address this, the relativistic reflection model {\tt relxillCp} \citep{garcia2014, dauser2014, jana2022broadband}, which represents the irradiation of the accretion disk by an {\tt nthcomp} Comptonization continuum, is incorporated into the model to account for the broad emission line instead of the Gaussian feature. Several parameters were fixed to ensure model stability and to reduce degeneracy. The spin (a) of the compact object was frozen at 0.1. \textbf{As the source is an eclipsing binary, the inclination angle ($\theta$) is expected to be high with previous studies suggesting values greater than $\sim 74.5^\circ$ \citep{intzand2003b}. Therefore, to minimize parameter degeneracy during spectral fitting, we fixed it to $60^\circ$. We also explored higher inclination values, however, the fit quality worsens and key parameters become poorly constrained.} The spectral fit favors a high Fe abundance ($A_{\rm Fe} \sim 10$), driven by the \texttt{relxillCp} component through the Fe line and Compton hump around $\sim 40$~keV; however, due to uncertainties in the \textit{LAXPC} background at these energies, we fix $A_{\rm Fe} = 5$, noting that adopting $A_{\rm Fe} = 1$ yields a slightly worse fit. {\bf The emissivity profile was modeled using a single emissivity index, which was fixed at the commonly adopted value of $q=3$, consistent with a standard disk illumination geometry used in several relativistic reflection studies \citep{Wilkins2012, Wilkins2018NS}.} The reflection fraction was set to -1, so that only the reflected component was produced by the model. The parameters of the Comptonization component, which represent the coronal properties, are tied to the corresponding values of {\tt relxillCp} model. The additional free parameters of the model are the normalization, the ionization parameter (log $\xi$) and the inner radius $R_{in}$. The 6.4 keV line and a broad hump at around 30 keV is well modeled with this component.

\begin{figure}
	\centering
	\includegraphics[width=0.73\linewidth]{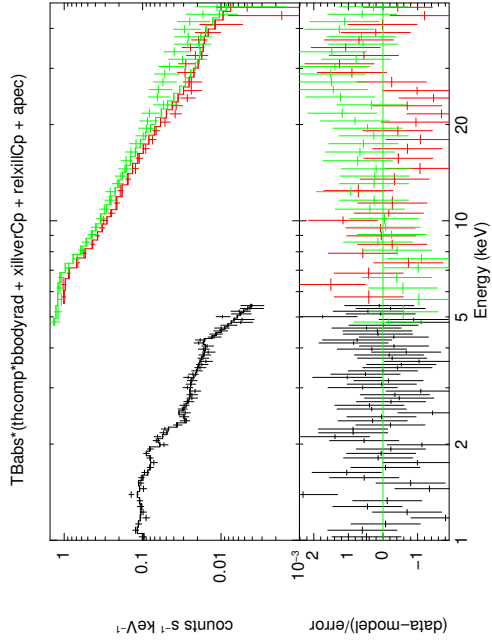}
	\caption{
		Folded energy spectra in the 1-50 keV band as observed by SXT (black),LAXPC20 (red), and LAXPC10 (green). The top panel shows the best-fitting spectra obtained with the model \texttt{constant*tbabs (thcomp*bbodyrad+xillverCp+relxillCp+apec)}, while the bottom panel shows the residuals.}
	\label{fig:2}
\end{figure}

We interpret the $\sim 4.2$ keV line as fluorescent emission from the surface of the neutron star. While reflection from the neutron star surface is expected to experience both gravitational redshift and rotational Doppler broadening, the present data do not require a relativistically broadened profile. The observed feature is spectrally narrow within the SXT resolution, and the signal-to-noise does not justify introducing additional relativistic broadening parameters. The feature is therefore modeled as a non-relativistic reflection component represented by the \textsc{Xspec} model {\tt xillverCp}, with the `redshift' parameter to be taken as the gravitational redshift at the surface of the neutron star. Thus, we use {\tt xillverCp} with reflection fraction set to -1, to get only the reflected component and tie the photon index and electron temperature to the ones in {\tt thcomp}. The free parameters are then the normalization, the ionization parameter, and the redshift, $z$. The best-fitting unfolded spectrum and residuals for this model are shown in Figure \ref{fig:1}(d), and the folded spectrum in counts space is shown in Figure \ref{fig:2}. The parameter values presented in Table \ref{tab:2} correspond to 2$\sigma$ confidence levels, and this error estimation is applied consistently throughout the paper. We note that the model fits the data well with $\chi^2$/d.o.f = 83.14/114 and the surface gravitational redshift is found to be z = 0.60$^{+0.03}_{-0.05}$. The absorbed flux in the 1.0-5.5 keV band is  $1.67^{+0.04}_{-0.07}\times10^{-11}$ erg s$^{-1}$ cm$^{-2}$.  Assuming a distance of 9.5 kpc \citep{kuulkers2003photospheric}, this corresponds  to a luminosity of $1.80^{+0.04}_{-0.08}\times10^{35}$ erg s$^{-1}$.  Using the alternative distance estimate of 6.7 kpc derived from  near-infrared observations of Terzan 6 \citep{valenti2007near}, the corresponding luminosity becomes 
$9.0^{+0.2}_{-0.4}\times10^{34}$ erg s$^{-1}$.


\subsection{Swift, Chandra \& XMM Newton}

 The \textit{Swift}, \textit{Chandra}, and \textit{XMM-Newton} observations considered here cover a wide range of flux levels. These data were selected to probe epochs outside the primary outburst peaks identified from long-term monitoring, rather than according to a strict quiescent luminosity criterion. In several cases, the \textit{Swift} flux exceeds that measured during the \textit{AstroSat} observation.
To check whether there is evidence for such a $\sim 4.2$ keV line in other data sets, we extracted spectra from \textit{Swift} (1.0-5.5) keV, \textit{Chandra} (1.0-5.5) keV and \textit{XMM-Newton} (1.0-10.0) keV from observations listed in Table \ref{tab:obs}. These energy ranges are narrower than the nominal instrumental ranges and were selected to exclude background-dominated regions and energy intervals with reduced calibration reliability. Since the exposure times of the {\it Swift} observations are significantly shorter and the count rates in the {\it Chandra} (1.0-5.5) keV and {\it XMM-Newton} spectra are low, we were unable to perform a detailed spectral analysis. Hence, we adopted the same spectral model as used for {\it AstroSat} (with a Gaussian line to represent the $4.2$ keV component) and allowed only for the interstellar absorption (\texttt{tbabs}) and the normalization parameters of all model components to vary. The $\sim 4.2$ keV line was not detected in the other observations, and we obtained the upper limits on the line flux and Equivalent width, which are listed in Table \ref{tab:3}.

\begin{figure}
	\centering
	\includegraphics[width=\linewidth]{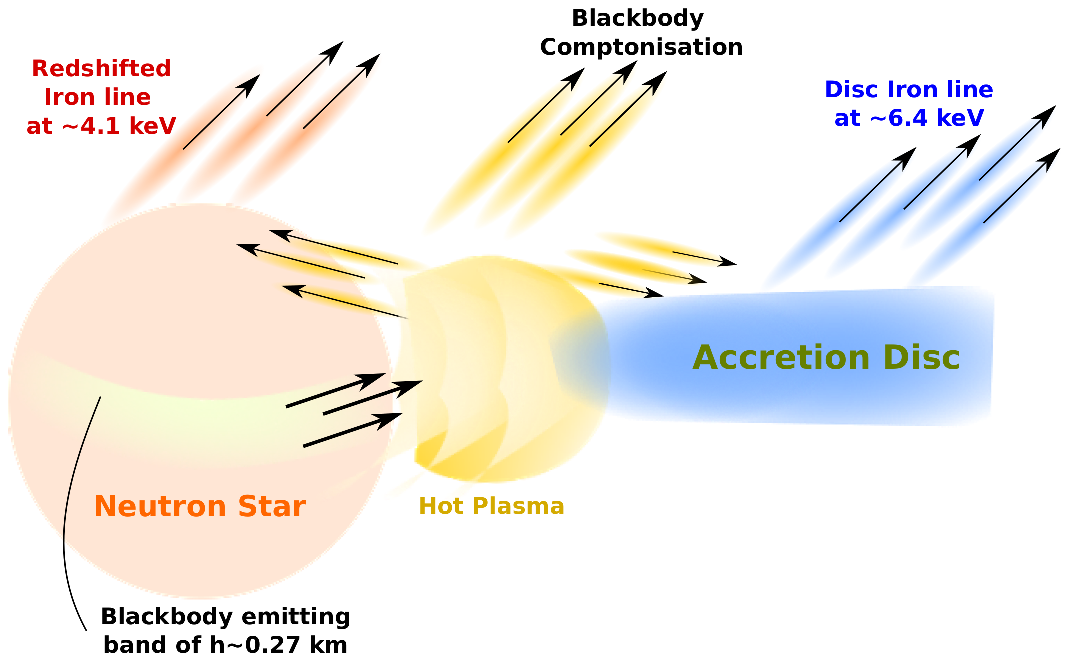}
	\caption{
		Schematic illustration of the proposed candidate emission and reflection geometry around the neutron star in GRS 1747$-$312. Thermal photons emitted from the neutron star surface are Comptonized in a hot corona, while a fraction of the radiation is reflected from the inner accretion disk, producing the broad Fe K$\alpha$ line at $\sim$6.4 keV. A portion of the reflected emission originates very close to the neutron star surface and experiences a strong gravitational redshift, giving rise to the observed candidate feature at 4.2 keV. The diagram is not to scale and is intended to qualitatively depict the spatial relationship between the neutron star, corona, and accretion disk.}
	\label{fig:grs_geometry}
\end{figure}

\begin{table}
	\centering
	\caption{Best-fitting parameters for the model
		\texttt{const*tbabs*(thcomp*bbodyrad+xillverCp+relxillCp+apec)} in the 1-5.5 keV band}
	\label{tab:2}
	\footnotesize
	\begin{tabular}{lll}
		\hline
		Component & Parameter & Value \\
		\hline
		\texttt{tbabs} & $n_{\mathrm{H}}\ (10^{22}\,\mathrm{cm}^{-2})$ & $1.06^{+0.09}_{-0.11}$ \\
		
		\texttt{thcomp} & $\Gamma_{\tau}$ & $1.72^{+0.08}_{-0.27}$ \\
		& $f_{\mathrm{cov}}$ & $0.85^{+0.11}_{-0.13}$ \\
		
		\texttt{bbodyrad} & $kT$ (keV) & $0.97^{+0.12}_{-0.14}$ \\
		& norm & $1.5^{+0.3}_{-0.5}$ \\
		
		\texttt{xillverCp} & $\log\xi$ & $3.0^{+0.5}_{-0.6}$ \\
		& $z$ & $0.61^{+0.03}_{-0.03}$ \\
		& norm ($10^{-5}$) & $3.8^{+2.4}_{-1.1}$ \\
		
		\texttt{relxillCp} & $R_{\mathrm{in}}\,(r_g)$ & $78^{+78}_{-50}$ \\
		& $\log\xi$ & $3.05^{+0.15}_{-0.23}$ \\
		& norm ($10^{-4}$) & $1.3^{+3.8}_{-3.8}$ \\
		
		\texttt{apec} & $kT$ (keV) & $0.96^{+0.07}_{-0.07}$ \\
		& $z\,(10^{-3})$ & $7^{+8}_{-13}$ \\
		& norm ($10^{-3}$) & $1.72^{+0.34}_{-0.53}$ \\
		
		\hline
		\texttt{Additional parameters} & norm ($10^{-5}$)$^{a}$ & $5.3^{+3.8}_{-2.7}$ \\
		& EqW (eV)$^{a}$ & $111^{+85}_{-57}$ \\
		& Flux$_{\rm abs}$ ($10^{-11}$erg cm$^{-2}$ s$^{-1}$) & $1.67^{+0.04}_{-0.07}$ \\
		& Flux$_{\rm unabs}$ ($10^{-11}$erg cm$^{-2}$ s$^{-1}$) & $2.53^{+0.23}_{-0.49}$ \\
		& Luminosity ($10^{35}$ erg s$^{-1}$)$^{b}$ & $1.8^{+0.4}_{-0.8}$ \\
		
		$\chi^2$/d.o.f. &  & $83.1/114$ \\
		\hline
	\end{tabular}
	
		$^{a}$Obtained from Gaussian fitting.\\
		$^{b}$Computed using the source distance.\\
\end{table}



\begin{table*}
	\centering
	\caption{Best-fitting values using model
		(\texttt{const*tbabs*(thcomp*bbodyrad+gaussian+relxillCp+apec)})
		for the \textit{AstroSat}, \textit{Swift}, \textit{Chandra}, and
		\textit{XMM-Newton} observations. The equivalent width (Eqwidth) is estimated at the upper limit of the LineE normalization. Fluxes and count rates are in the 1-5.5 keV energy range. The listed observations correspond to epochs outside major outburst peaks and were selected based on adequate signal-to-noise for constraining the 4.2 keV line.}
	\label{tab:3}
	\footnotesize
	\begin{tabular}{ccccccc}
		\hline
		Telescope & Observation ID &  Count rate & Flux & Eqwidth & LineE norm & $\chi^2$/d.o.f. \\
		&                &  (s$^{-1}$) & ($10^{-11}$erg cm$^{-2}$ s$^{-1}$) & (eV) & ($10^{-5}$) &  \\
		\hline
		\textit{AstroSat} & G07\_020T01\_9000001398 &  0.218 & $1.72^{+0.03}_{-0.03}$ & $111^{+85}_{-57}$ & $5.3^{+3.8}_{-2.7}$ & 87.3/120 \\
		\hline
		\multirow{8}{*}{\textit{Swift}}
		& 00032761023 & 1.05 & $47.6^{+2.6}_{-2.8}$ & $<167$ & $<360$ & 27.5/20 \\
		& 00032761002 & 0.57 & $10.8^{+0.8}_{-0.9}$ & $<110$ & $<52$  & 18.7/16 \\
		& 00032761039 & 0.80 & $26.9^{+1.7}_{-1.8}$ & $<102$ & $<132$ & 20.2/17 \\
		& 00032761004 &  0.31 & $8.2^{+0.6}_{-0.7}$  & $<83$  & $<31$  & 17.6/17 \\
		& 00032761007 & 0.29 & $7.7^{+0.6}_{-0.7}$  & $<69$  & $<23$  & 18.6/14 \\
		& 00032761001 &  0.63 & $35.0^{+2.3}_{-2.6}$ & $<69$  & $<109$ & 15.1/19 \\
		& 00032761008 &  0.96 & $49.4^{+2.7}_{-2.8}$ & $<59$  & $<144$ & 24.4/20 \\
		& 00032761009 &  0.63 & $38.7^{+1.9}_{-2.2}$ & $<30$  & $<53$  & 28.0/24 \\
		\hline
		\multirow{5}{*}{\textit{Chandra}}
		& 26514 & 0.13 & $0.190^{+0.006}_{-0.007}$ & $<104$ & $<1.62$ & 12.7/11 \\
		& 23443 & 0.03 & $0.050^{+0.002}_{-0.002}$ & $<98$  & $<0.38$ & 15.5/10 \\
		& 23441 &   0.06 & $0.070^{+0.006}_{-0.006}$ & $<41$  & $<0.27$ & 11.7/10 \\
		& 26515 &  0.11 & $0.180^{+0.006}_{-0.006}$ & $<41$  & $<0.50$ & 11.5/10 \\
		& 26513 &  0.08 & $0.110^{+0.005}_{-0.005}$ & $<18$  & $<0.37$ & 11.8/10 \\
		\hline
		\textit{XMM-Newton} & 0206990101 &  0.02 & $0.009^{+0.001}_{-0.001}$ & $<286$ & $<0.04$ & 108.2/74 \\
		\hline
	\end{tabular}
\end{table*}
\section{Discussion \texorpdfstring{\&}{&} Conclusion}\label{sec:discussion}
While narrow absorption features at $\sim 4$ keV have been reported in the spectra of Type 1 X-ray bursts from neutron star systems \citep{waki1985discovery, inoue1985discovery, magnier19894, nakamura1988detection}, to our knowledge, this is the first report of a candidate emission feature at this energy. The luminosity of the source during {\it AstroSat} observation was around $\sim (0.9-1.80) \times 10^{35}$ erg s$^{-1}$, which suggests that the emission occurs for a particular geometry which is perhaps realized only at such sub-Eddington luminosities.

 Given the crowded nature of the field and the relatively large extraction region, contamination from nearby sources, in particular Terzan 6 X-2, cannot be fully excluded. During quiescence, {\it Chandra} has resolved Terzan 6 X-2 (located $\sim$ 1 arcsec from GRS 1747$-$312) as a distinct source, identified as an eclipsing LMXB, that is known to undergo outbursts. Although there is currently no direct evidence that Terzan 6 X-2 was active during our observations, it can reach luminosities comparable to a significant fraction of that reported in \citet{van2024discovery}, and may therefore contribute to the observed spectrum without dominating the emission.

In principle, some of the {\it Swift} and in particular the {\it AstroSat} observations could have been obtained during an outburst of Terzan 6 X-2 rather than GRS 1747$-$312. However, it appears unlikely that both sources were simultaneously in outburst with nearly identical flux levels during the {\it AstroSat} observation. Therefore, while treating the emission as originating predominantly from a single source remains a reasonable working assumption, we explicitly acknowledge that contamination from Terzan 6 X-2 cannot be ruled out. Consequently, the spectral features reported in this work, including the $\sim$ 4.2 keV feature, should be regarded as tentative.
 
The spectral fitting provides clues to the geometry of the system during this state. The blackbody normalization of $\sim 1.5$ yields a radius of $\sim$ 1.16 km for a distance of $\sim 9.5$ kpc, significantly smaller than the expected neutron star radius. In other words, the inferred surface area of the black body emitter, $\sim 4 \pi (1.07)^2$ km$^2$ is smaller than the total surface area of the neutron star. This suggests that the emission may be coming from a strip along the equator of the neutron star with a surface area equal to $2 \pi R_{\text{NS}} 2W$, where $W$ is the half-width of the strip. For a neutron star radius of $R_{\text{NS}} \sim 10$ kms, this implies a width of $\sim 0.27$ kms. This suggests a possible geometry which is schematically represented in Figure \ref{fig:grs_geometry}. Thermal photons emitted from the neutron star surface are Comptonized by a hot corona, with part of the emission reflected by the inner accretion disk, producing the observed broad Fe K$_\alpha (\sim 6.4$ keV) line. A fraction of the photons are also reflected from the surface of the neutron star giving rise to the candidate red-shifted Iron line ($\sim 4.2$ keV). We note that emission lines from lighter elements such as Mg, Si, and Ar are expected in the $\sim$ 2$-$4 keV band. However, the observed feature at $\sim 4.2$ keV is difficult to reconcile with these species. The strongest transitions of Mg, Si, and Ar occur at significantly lower energies ($\sim 1.3$-$3.3$ keV), and invoking these ions would require large blueshifts ($\gtrsim 20$ - $40\%$), which are not expected and would likely produce additional shifted features that are not observed. We therefore consider an origin in lighter elements to be unlikely.

Perhaps, for systems with higher accretion rates, the corona may expand and cover a larger fraction of the neutron star surface, thereby suppressing or preventing the detection of the 4.2 keV emission line. In addition, one would need to speculate that when the source is in quiescence (i.e. at lower luminosity) again the geometry may be that of an extended corona leading to potentially weaker reflection signatures. However, a detailed analysis needs to be undertaken to understand the difference in geometry and physical conditions on the surface of the neutron star, which causes the candidate line emission to remain undetected at high luminosities. The gravitational redshift at the surface of a neutron star can be approximated to be  \citep[e.g.][]{Glendenning:1997wn},
\begin{equation} \label{eq:1}
	z = \frac{1}{\sqrt{1 - \frac{2GM}{Rc^2}}} - 1 
\end{equation}
Rewriting this expression, the neutron star radius can be expressed as a function of mass and redshift:
	\[
	R = \frac{2GM}{c^2\left[1 - (1+z)^{-2}\right]}.
	\]
	
	For $z \sim 0.6$, this yields
	\[
	R \simeq 6.8 \left(\frac{M}{1.4\,M_\odot}\right)\ \mathrm{km},
	\]
	explicitly showing the dependence on the assumed neutron star mass.

We note that there are no obvious atomic transitions at $\sim 4.2$ keV, which may lead to an instrumental feature. The {\tt apec} model fitting to the data set (with any redshift constrained to be $< 0.01$) shows that the energy calibration of the SXT is adequate. Moreover, such a line has not been reported for any of the several SXT observations of other sources. The line can be confirmed with a long {\it Swift}/XRT observation or any observation by {\it XMM-Newton, Chandra, Nustar} or {\it XRISM}. The challenge is to catch the source at a flux level of $\sim 10^{-11}$ erg s$^{-1}$ cm$^{-2}$. All-sky monitors such as MAXI are generally not sensitive enough to detect such low-level activity. However, the recently launched Einstein Probe mission may provide sufficient sensitivity to detect similar faint outbursts and enable deeper follow-up observations \citep{yuan2025}. Such observations would be important to probe features such as the 4.2 keV emission line, if present. So, continuous monitoring with sensitive X-ray observatories in future would therefore help confirm the presence of the 4.2 keV feature at these low flux levels.

\section*{Acknowledgements}
The authors thank the anonymous reviewer for constructive comments to improve the manuscript. This work has used the data from the Soft X-ray Telescope (SXT) developed at TIFR, Mumbai, and the SXT POC at TIFR is thanked for verifying and releasing the data via the ISSDC data archive and providing the necessary software tools. We also acknowledge the support of the LAXPC POC at TIFR, as well as the AstroSat Science Support Cell at IUCAA, for making the data and software available. This work was supported by the Inter-University Centre for Astronomy and Astrophysics (IUCAA), Pune. A.L. Chanu thank the IUCAA staff for their support and hospitality during the course of this research. A.L. Chanu  also acknowledge the use of archival data obtained through the AstroSat Science Support Cell and other mission archives.



\bibliographystyle{elsarticle-harv} 
\bibliography{references}

@article{Barbuy1997ntt,
  author = {Barbuy, B. and Ortolani, S. and Bica, E.},
  title = {NTT V, I, z photometry of the metal-rich bulge globular cluster Terzan 6},
  journal = {A\&AS},
  volume = {122},
  pages = {483},
  year = {1997}
}

@article{hulleman2003bursts,
  author = {Hulleman, F. and others},
  title = {Bursts, eclipses, dips and a refined position for the luminous low-mass X-ray binary in the globular cluster Terzan 6},
  journal = {A\&A},
  volume = {406},
  pages = {233},
  year = {2003}
}

@article{kuulkers2003photospheric,
  author = {Kuulkers, E. and others},
  title = {Photospheric radius expansion X-ray bursts as standard candles},
  journal = {A\&A},
  volume = {399},
  pages = {663},
  year = {2003}
}

@article{intzand2005outburst,
  author = {in't Zand, J. J. M. and others},
  title = {The outburst behavior of the X-ray burster GRS 1747–312 in Terzan 6},
  journal = {A\&A},
  volume = {441},
  number = {2},
  pages = {675--682},
  year = {2005}
}

@article{in2004bepposax,
  author = {in’t Zand, J. J. M. and others},
  title = {BeppoSAX-WFC monitoring of the Galactic Center region},
  journal = {Nucl. Phys. B Proc. Suppl.},
  volume = {132},
  pages = {486},
  year = {2004}
}

@article{predehl1991rosat,
  author = {Predehl, P. and Hasinger, G. and Verbunt, F.},
  title = {ROSAT discovery of bright X-ray sources in globular cluster Terzan 6 and NGC 6652},
  journal = {A\&A},
  volume = {246},
  pages = {L21},
  year = {1991}
}

@article{pavlinsky1994x,
  author = {Pavlinsky, M. N. and Grebenev, S. A. and Sunyaev, R. A.},
  title = {X-ray images of the Galactic Center obtained with ART-P/GRANAT: Discovery of new sources, variability of persistent sources, and localization of X-ray bursters},
  journal = {ApJ},
  volume = {425},
  pages = {110},
  year = {1994}
}

@article{iwai2017broad,
  author = {Iwai, M. and others},
  title = {A broad spectral feature detected during the cooling phase of a type I X-ray burst from GRS 1747-312 with Suzaku},
  journal = {PASJ},
  volume = {69},
  pages = {61},
  year = {2017}
}

@article{bazzano2000orbital,
  author = {Bazzano, A. and others},
  title = {The orbital period of the recurrent X-ray transient in Terzan 6},
  journal = {A\&A},
  volume = {355},
  pages = {145--154},
  year = {2000}
}

@article{saji2016peculiar,
  author = {Saji, K. and others},
  title = {Peculiar lapse of periodic eclipsing event at low-mass X-ray binary GRS 1747-312 during Suzaku observation in 2009},
  journal = {PASJ},
  volume = {68},
  pages = {1},
  year = {2016}
}

@article{painter2024possible,
  author = {Painter, C. and others},
  title = {A possible third body in the X-ray system GRS 1747-312 and models with higher-order multiplicity},
  journal = {MNRAS},
  volume = {529},
  pages = {245},
  year = {2024}
}

@article{vats2018quiescent,
  author = {Vats, S. and others},
  title = {The quiescent state of the neutron-star X-ray transient GRS 1747-312 in the globular cluster Terzan 6},
  journal = {MNRAS},
  volume = {477},
  pages = {2494},
  year = {2018}
}

@InCollection{Bahramian2024,
  author    = {Bahramian, Arash and Degenaar, Nathalie},
  title     = {Low-Mass X-ray Binaries},
  booktitle = {Handbook of X-ray and Gamma-ray Astrophysics},
  publisher = {Springer Nature Singapore},
  address   = {Singapore},
  year      = {2024},
  pages     = {3657--3718},
  doi       = {10.1007/978-981-19-6960-7_94}
}

@article{revnivtsev2002localization,
  author = {Revnivtsev, M. G. and Trudolyubov, S. P. and Borozdin, K. N.},
  title = {Localization of X-ray sources in six Galactic globular clusters from Chandra data},
  journal = {Astronomy Letters},
  volume = {28},
  pages = {237},
  year = {2002}
}

@article{van2024discovery,
  author = {Van den Berg, M. and others},
  title = {Discovery of a Second Eclipsing, Bursting Neutron Star Low-mass X-Ray Binary in the Globular Cluster Terzan 6},
  journal = {ApJ},
  volume = {966},
  pages = {217},
  year = {2024}
}

@article{agrawal2006broad,
  author = {Agrawal, P. C.},
  title = {A broad spectral band Indian Astronomy satellite ‘Astrosat’},
  journal = {Advances in Space Research},
  volume = {38},
  pages = {2989},
  year = {2006}
}

@article{agrawal2017large,
  author = {Agrawal, P. C. and others},
  title = {Large area X-ray proportional counter (LAXPC) instrument on AstroSat and some preliminary results from its performance in the orbit},
  journal = {J. Astrophys. Astron.},
  volume = {38},
  pages = {1},
  year = {2017}
}

@article{yadav2016astrosat,
  author = {Yadav, J. S. and others},
  title = {AstroSat/LAXPC reveals the high-energy variability of GRS 1915+105 in the $\chi$ class},
  journal = {ApJ},
  volume = {833},
  pages = {27},
  year = {2016}
}

@article{antia2017calibration,
  author = {Antia, H. M. and others},
  title = {Calibration of the large area X-ray proportional counter (LAXPC) instrument on board AstroSat},
  journal = {ApJS},
  volume = {231},
  pages = {10},
  year = {2017}
}

@article{antia2021large,
  author = {Antia, H. M. and others},
  title = {Large Area X-ray Proportional Counter (LAXPC) in orbit performance: Calibration, background, analysis software},
  journal = {J. Astrophys. Astron.},
  volume = {42},
  pages = {32},
  year = {2021}
}

@article{beri2019thermonuclear,
  author = {Beri, A. and others},
  title = {Thermonuclear X-ray bursts in rapid succession in 4U 1636--536 with AstroSat-LAXPC},
  journal = {MNRAS},
  volume = {482},
  pages = {4397},
  year = {2019}
}

@article{nath2022astrosat,
  author = {Nath, A. and others},
  title = {AstroSat observation of rapid type-I thermonuclear burst from low-mass X-ray binary GX 3+1},
  journal = {J. Astrophys. Astron.},
  volume = {43},
  pages = {93},
  year = {2022}
}

@article{singh2017soft,
  author = {Singh, K. P. and others},
  title = {Soft X-ray focusing telescope aboard AstroSat: design, characteristics, and performance},
  journal = {J. Astrophys. Astron.},
  volume = {38},
  pages = {1},
  year = {2017}
}

@article{kaastra2016optimal,
  author = {Kaastra, J. S. and Bleeker, J. A. M.},
  title = {Optimal binning of X-ray spectra and response matrix design},
  journal = {A\&A},
  volume = {587},
  pages = {A151},
  year = {2016}
}

@article{struder2001european,
  author = {Strüder, L. and others},
  title = {The European photon imaging camera on XMM-Newton: the pn-CCD camera},
  journal = {A\&A},
  volume = {365},
  pages = {L18},
  year = {2001}
}

@article{turner2001european,
  author = {Turner, M. J. L. and others},
  title = {The European photon imaging camera on XMM-Newton: the MOS cameras},
  journal = {A\&A},
  volume = {365},
  pages = {L27},
  year = {2001}
}

@article{zdziarski2020spectral,
  author = {Zdziarski, A. A. and others},
  title = {Spectral and temporal properties of Compton scattering by mildly relativistic thermal electrons},
  journal = {MNRAS},
  volume = {492},
  pages = {5234},
  year = {2020}
}

@article{wilms2000absorption,
  author  = {Wilms, J. and Allen, A. and McCray, R.},
  title   = {On the Absorption of X-Rays in the Interstellar Medium},
  journal = {ApJ},
  volume  = {542},
  number  = {2},
  pages   = {914--924},
  year    = {2000},
  doi     = {10.1086/317016}
}

@article{jana2022broadband,
  author = {Jana, A. and Chang, H.-K. and Chatterjee, A. and Naik, S. and Safi-Harb, S.},
  title = {Broadband X-Ray Spectroscopy and Estimation of Spin of the Galactic Black Hole Candidate GRS 1758--258},
  journal = {ApJ},
  volume = {936},
  pages = {3},
  year = {2022}
}

@article{garcia2014,
  author = {Garcia, J. and others},
  title = {Improved reflection models of black hole accretion disks: Treating the angular distribution of X-rays},
  journal = {ApJ},
  volume = {782},
  pages = {76},
  year = {2014}
}

@article{dauser2014,
  author = {Dauser, T. and García, J. and Parker, M. L. and Fabian, A. C. and Wilms, J.},
  title = {The role of the reflection fraction in constraining black hole spin},
  journal = {MNRAS Letters},
  volume = {444},
  pages = {L100},
  year = {2014}
}

@book{Glendenning:1997wn,
    author = "Glendenning, N. K.",
    title = "{Compact stars: Nuclear physics, particle physics, and general relativity}",
    year = "1997"
}

@article{waki1985discovery,
  title={Discovery of absorption lines in X-ray burst spectra from X1636-536},
  author={Waki, Izumi and Inoue, Hajime and Koyama, Katsuji and Matsuoka, Masaru and Murakami, Toshio and Ogawara, Yoshiaki and Ohashi, Takaya and Tanaka, Yasuo and Hayakawa, Satio and Tawara, Yuzuru and others},
  journal={Publications of the Astronomical Society of Japan},
  volume={36},
  number={4},
  pages={819--830},
  year={1985},
  publisher={Oxford University Press}
}

@article{inoue1985discovery,
  title={Discovery of absorption lines in X-ray burst spectra from X1636-536.},
  author={Inoue, H and Koyama, K},
  journal={Advances in Space Research},
  volume={5},
  number={3},
  pages={91--94},
  year={1985}
}

@article{magnier19894,
  title={A 4.1 keV spectral feature in a type 1 X-ray burst from EXO 1747--214},
  author={Magnier, Eugene and Lewin, Walter HG and van Paradijs, Jan and Tan, Jianmin and Penninx, Wim and Damen, Eugene},
  journal={Monthly Notices of the Royal Astronomical Society},
  volume={237},
  number={3},
  pages={729--738},
  year={1989},
  publisher={Oxford University Press Oxford, UK}
}

@article{nakamura1988detection,
  title={Detection of absorption lines in the spectra of x-ray bursts from X 1608-52},
  author={Nakamura, Norio and Inoue, Hajime and Tanaka, Yasuo},
  journal={Publications of the Astronomical Society of Japan},
  volume={40},
  number={2},
  pages={209--217},
  year={1988},
  publisher={Oxford University Press}
}

@article{valenti2007near,
  title={Near-infrared properties of 24 globular clusters in the galactic bulge},
  author={Valenti, Elena and Ferraro, FRANCESCO ROSARIO and Origlia, L},
  journal={The Astronomical Journal},
  volume={133},
  number={4},
  pages={1287--1301},
  year={2007}
}

@article{misra2017astrosat,
  author = {Misra, R. and others},
  title = {AstroSat/LAXPC instrument and background modeling},
  journal = {Journal of Astrophysics and Astronomy},
  volume = {38},
  pages = {30},
  year = {2017},
  doi = {10.1007/s12036-017-9451-6}
}

@article{disalvo2023review,
  author = {Di Salvo, Tomaso and others},
  title = {Neutron Star Low-Mass X-ray Binaries},
  journal = {Astrophysics and Space Science Library},
  volume = {461},
  pages = {153--195},
  year = {2023},
  doi = {10.1007/978-981-99-1591-5_4}
}

@article{intzand2003b,
  author = {in't Zand, J. J. M. and Verbunt, F. and Heise, J. and Markwardt, C. B.},
  title = {The eclipsing and bursting low-mass X-ray binary GRS 1747-312 in Terzan 6},
  journal = {Astronomy \& Astrophysics},
  volume = {406},
  pages = {233--244},
  year = {2003},
  doi = {10.1051/0004-6361:20030764}
}

@article{Wilkins2012,
  author  = {Wilkins, D. R. and Fabian, A. C.},
  title   = {Understanding X-ray reflection emissivity profiles in AGN: locating the X-ray source},
  journal = {MNRAS},
  volume  = {424},
  pages   = {1284--1296},
  year    = {2012},
  doi     = {10.1111/j.1365-2966.2012.21308.x}
}

@article{Wilkins2018NS,
  author       = {Wilkins, D. R.},
  title        = {On the illumination of neutron star accretion discs},
  journal      = {Monthly Notices of the Royal Astronomical Society},
  volume       = {475},
  number       = {1},
  pages        = {748--756},
  year         = {2018},
  doi          = {10.1093/mnras/stx3169}
}

@article{yuan2025,
  author = {Yuan, Weimin and Dai, Lixin and Feng, Hua and Jin, Chichuan and Jonker, Peter and Kuulkers, Erik and others},
  title = {Science objectives of the Einstein Probe mission},
  journal = {Science China Physics, Mechanics \& Astronomy},
  year = {2025},
  volume = {68},
  number = {3},
  pages = {239511},
  doi = {10.1007/s11433-024-2600-3}
}





\end{document}